\documentclass[11pt,letterpaper]{article}
\usepackage{epsfig}
\usepackage{graphicx}
\usepackage{authblk}
\usepackage[numbers]{natbib}

\title{Supporting Developers in Porting Software via Combined Textual
  and Structural Analysis of Software Artifacts}

\author[1]{Kostadin Damevski}
\author[2]{David Shepherd}
\author[2]{Nicholas Kraft}
\author[3]{Lori Pollock}
\affil[1]{Department of Computer Science, Virginia Commonwealth University}
\affil[2]{ABB Corporate Research, Raleigh, NC}
\affil[3]{Department of Computer Science, University of Delaware}

\date{\vspace{-5ex}}

\begin{document}
\maketitle
\thispagestyle{empty}


In the engineering and scientific domains software commonly has a long
lifespan, lasting decades instead of years. Due to this lifespan,
software often outlives the current generation of hardware, and in
turn needs to be modified to execute on newer classes of hardware
architectures~\cite{carver_software_2007}. Supporting developers in
this difficult software maintenance activity is very important in
order to improve their productivity, reduce bugs, and prevent
architectural erosion ~\cite{faulk_scientific_2009}.

Software engineering researchers have found that developers commonly
begin maintenance tasks by first identifying and then comprehending
the relevant program units for that
task~\cite{ko_exploratory_2006}. Program comprehension at the outset
of maintenance tasks has been well studied and supported by a number
of tools, utilizing static and dynamic information extracted from the
code base. Research in the last decade in applying textual analysis
approaches to source code, adapted from the fields of natural language
processing and information retrieval, has produced notable results and
resulted in the proliferation of program comprehension tools based on
this type of information, often combined with static or dynamic
sources~\cite{dit_feature_2013}. Textual approaches have now matured
to be applicable to the problem of enabling long-lasting high
performance scientific and engineering software, a domain where
natural language information is usually more challenging to extract
than in general purpose software.

Based on these opportunities, our position is that {\bf the use of
  textual and structural analyses coupled with platform usage patterns
  extracted by mining similar open source projects can improve the
  quality and efficiency of scientific and engineering software
  maintenance.}  Our position is based on two key insights: 1)
combining structural and textual analyses can reveal information to
support scientific and engineering software maintenance; 2) mining
software repositories of projects that exist within the same platform
can reveal patterns that could aid developers in identifying and
comprehending relevant program elements for porting. We envision that
research in these directions could yield software comprehension and
recommendation tools that could be applied to a code base in order to
improve porting to new platforms during maintenance. In the following
we discuss each of the two key ideas in turn.

\noindent
{\bf Combining structural and textual analyses.}  Natural language
processing of identifiers and comments embedded in source code has
become recognized as a necessary tool in aiding developers accomplish
certain software engineering tasks, such as feature
location~\cite{dit_feature_2013}, software
modularization~\cite{garcia_comparative_2013}, and automated code
summarization~\cite{moreno_automatic_2013}. Numerous text analysis
algorithms and approaches, including topic modeling, text retrieval,
and others have been found to produce meaningful results on source
code terms, when configured and adapted to this type of textual
information.  Such approaches that adapt text analysis to software
often include the use of abbreviation expansion techniques (e.g. attr
$\rightarrow$ attribute, param $\rightarrow$ parameter, println
$\rightarrow$ print line), the use of identifier splitting techniques
(e.g. sortedList $\rightarrow$ sorted list, textbox $\rightarrow$ text
box), as well as the construction of various dictionaries (e.g. for
synonyms) containing words from this context.  Certain kinds of text
analyses are heavily reliant on the quality of names in the source
code, while other kinds (e.g., topic modeling, statistical models) are
reliant only on the consistency of names (e.g., to relate artifacts
based on similar term co-occurrences), which makes them able to
overcome certain types of naming deficiencies in source code.

While natural language text is commonly organized in documents,
software offers a much richer set of structural organization,
including class inheritance hierarchy, method call graph, control flow
graphs, class (or file) dependencies, and data flow graphs. In working
with natural language in the context of software engineering, one has
to choose a structural granularity at which to process text. Instead
of just as a means of decomposition, using the program structure as
another source of information to the embedded natural language text
has recently produced improved results on many software engineering
applications.

{\em While complex scientific programs can be written in low level
languages, and without much documentation, textual analysis for source
code is likely mature enough to overcome these challenges.  Coupled
with statically-extracted structural information, it should yield a
wealth of information to be presented to developers and aid porting to
a new platform.}

\noindent
{\bf Mining software repositories of projects that execute in similar
  environment.}  In interacting with hardware resources, applications
tend to follow a fixed set of patterns, exhibited by the API calls and
other program structure, as well as by the natural language terms
embedded in that portion of the source code. By mining existing
examples of applications on the same platform to learn how hardware
resources are accessed, a tool could help developers by identifying
resource-relevant source code and, via its natural language
representation, characterizing that code's intent. Many repositories
containing such information in their histories are publicly available,
e.g. as part of the more than 20 million repositories hosted on
GitHub.
 
{\em Research in approaches that mine and distill hardware environment
  information from source code repositories can be important in the
  construction of recommendation systems for developers in porting
  code to new platforms.}

%

\bibliographystyle{plain}

\end{document}